\documentclass[aps,prl,twocolumn,showpacs,showkeys,footinbib,amsmath,amssymb,superscriptaddress]{revtex4-2}

\usepackage{amsmath}
\usepackage{amssymb}
\usepackage{graphicx}
\usepackage{bm}
\usepackage{color}
\usepackage{relsize}
\usepackage{braket}
\usepackage{bbold}
\usepackage{txfonts}
\usepackage{epstopdf}

\usepackage[colorlinks,
linkcolor=blue,
anchorcolor=blue,
citecolor=magenta,
]{hyperref}

\begin{document}	
\title{Two-Dimensional Ferroelectric Altermagnets: From Model to Material Realization}

\author{Ziye Zhu}
\affiliation{Eastern Institute for Advanced Study, Eastern Institute of Technology, Ningbo, Zhejiang 315200, China}
\affiliation{International Center for Quantum Design of Functional Materials (ICQD), and Hefei National Laboratory, University of Science and Technology of China, Hefei, 230026, China}

\author{Xunkai Duan}
\affiliation{Eastern Institute for Advanced Study, Eastern Institute of Technology, Ningbo, Zhejiang 315200, China}
\affiliation{School of Physics and Astronomy, Shanghai Jiao Tong University, Shanghai 200240, China}

\author{Jiayong Zhang}
\affiliation{Eastern Institute for Advanced Study, Eastern Institute of Technology, Ningbo, Zhejiang 315200, China}
\affiliation{International Center for Quantum Design of Functional Materials (ICQD), and Hefei National Laboratory, University of Science and Technology of China, Hefei, 230026, China}
\affiliation{School of Physical Science and Technology, Suzhou University of Science and Technology, Suzhou, 215009, China}

\author{Bowen Hao}
\affiliation{Eastern Institute for Advanced Study, Eastern Institute of Technology, Ningbo, Zhejiang 315200, China}

\author{Igor \v{Z}uti\'c}
\affiliation{Department of Physics, University at Buffalo, State University of New York, Buffalo, New York 14260, USA}

\author{Tong Zhou}
\email{tzhou@eitech.edu.cn}
\affiliation{Eastern Institute for Advanced Study, Eastern Institute of Technology, Ningbo, Zhejiang 315200, China}


\date{\today}

\begin{abstract}
	Multiferroic altermagnets offer new opportunities for magnetoelectric coupling and electrically tunable spintronics. However, due to intrinsic symmetry conflicts between altermagnetism and ferroelectricity, achieving their coexistence, known as ferroelectric altermagnets (FEAM), remains an outstanding challenge, especially in two-dimensional (2D) systems. Here, we propose a universal, symmetry-based design principle for 2D FEAM, supported by tight-binding models and first-principles calculations. We show that lattice distortions can break spin equivalence and introduce the necessary rotation-related symmetry, enabling altermagnetism with electrically reversible spin splitting. Guided by this framework, we identify a family of 2D vanadium oxyhalides and sulfide halides as promising FEAM candidates. In these compounds, pseudo Jahn-Teller distortions and Peierls-like dimerization cooperatively establish the required symmetry conditions. We further propose the magneto-optical Kerr effect as an experimental probe to confirm FEAM and its electric spin reversal. Our findings provide a practical framework for 2D FEAM and advancing electrically controlled spintronic devices.
\end{abstract}
	
	\maketitle

Electric control of magnetism remains a longstanding goal in multiferroics and spintronics, with the potential to revolutionize high-density data storage and energy-efficient spintronic devices~\cite{zutic2004:RMP,Eerenstein2006,Cheong2007,Dong2015}. Multiferroic materials, which exhibit coexisting magnetic and (anti)ferroelectric orders, are particularly enticing for this purpose because they couple electric and magnetic properties within a single phase. Recent discoveries of altermagnets (AM)~\cite{Smejkal2022a,Vsmejkal2022:beyond,Bai2024,Wei2024,Cheong2024Alter,Fender2025,song2025altermagnets, Yuan2020:PRB,
	krempasky2024altermagnetic,SongAMNature2025,jiang2025metallic,
	zhu2023multipiezo, JunweiliuNC2021, Guo2023:quantum,Guo2024valleyAM_PRB,wang2024Electric,  Yan2024PRL_Skyrmion,zhang2024:predictable,camerano2025multiferroic,Samanta2025Spin,qian2025fragile}, a new class of collinear magnets that display spin splitting without net magnetization, further expand this research frontier. Integrating (anti)ferroelectricity with altermagnetism offers a pathway to fast, reversible control of magnetism and spin polarization~\cite{Duan2025PRL, Gu2025PRL, Smejkal2024arxiv,wright2025altermagnets,Sun2024a,MenghaoWu_slidingAM,cao2024designing}.

A key challenge in realizing such multiferroic altermagnets lies in establishing the appropriate symmetry, particularly rotation-related symmetry, $R$, that connects opposite-spin sublattices~\cite{Vsmejkal2022:beyond,Duan2025PRL}. While conventional ferroelectricity often preserves translation symmetry, $t$, leading to traditional antiferromagnets (AFM), antiferroelectricity naturally introduces the $R$  essential for AM~\cite{Duan2025PRL}. This has recently led to the concept of antiferroelectric altermagnets, where an electric field toggles spin polarization on and off by switching between antiferroelectric and ferroelectric phases~\cite{Duan2025PRL}. While antiferroelectric altermagnets have been realized in both three-dimensional (3D) perovskites and two-dimensional (2D) van der Waals materials, ferroelectric altermagnets (FEAM) are exceedingly rare, with only two switchable 3D FEAM identified across the entire MAGNDATA database~\cite{Gu2025PRL}.

Despite their promise, FEAM research has so far been limited to 3D bulk crystals~\cite{Gu2025PRL, Smejkal2024arxiv}. Extending FEAM to 2D systems is both timely and essential for the development of nanoscale devices, where atomically thin materials offer key advantages such as improved scalability, tunable physical properties, and facile integration with flexible electronics~\cite{bhimanapati2015recent}. However, 2D FEAM remains largely unexplored, and a general theoretical framework that explains how ferroelectricity can induce altermagnetism in 2D systems is still lacking. This absence not only limits the fundamental understanding of FEAM but also hinders systematic materials discovery. Given the rising demand for ultrathin materials with electrically switchable magnetism, identifying and designing 2D FEAM has become an urgent goal.

In this work, we overcome these challenges by introducing a universal design principle for 2D FEAM and confirming its feasibility through both model analysis and first-principles calculations. We first construct a tight-binding (TB) model showing how lattice distortions break translation symmetry to enable spin‐inequivalent hopping, triggering altermagnetism. Crucially, reversing the FE polarization flips the spin splitting by interchanging sublattice‐dependent hoppings. Guided by this model, we identify a family of well-established 2D materials, vanadium oxyhalides and sulfide halides (VOX$_2$ and VSX$_2$, with X = Cl, Br, I)~\cite{tan2019two, ai2019intrinsic, xu2020electric, zhang2021peierls}, as promising FEAM candidates. In these compounds, ferroelectricity arises from pseudo Jahn-Teller distortions, while Peierls-like dimerization induces the $R$ symmetry to realize the FEAM. Our simulations confirm the reversibility of the ferroelectrically controlled spin splitting, and we propose the magneto‐optical Kerr effect (MOKE) as a viable experimental probe. By integrating symmetry analysis, effective model, and first‐principles validation, our work not only establishes a promising pathway for achieving 2D FEAM but also lays the groundwork for their future experimental realization and spintronic applications.

\begin{figure*}[t!]
	\centering
	\vspace{0.2cm}
	\includegraphics*[width=0.8\textwidth]{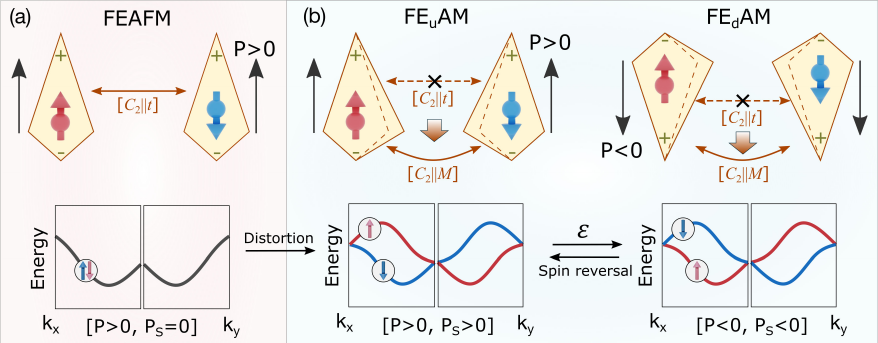}
	\vspace{-0.2cm}
	\caption{ Design principle for FEAM. Magnetic atoms with opposite spins (red and blue arrows) form AFM sublattices, while parallel local electric polarizations (black arrows), arising from asymmetric charge distributions, produce a FE state with nonzero net electric polarization, $\bf P$. \textbf{(a)} FEAFM: Magnetic sublattices are directly connected by translation symmetry, $t$, resulting in conventional AFM with a vanishing spin splitting, ${P_S}$ = 0. \textbf{(b)} FEAM: Lattice distortions break $t$ but preserve mirror symmetry, $M$, between magnetic sublattices, creating AM with $\bf {P_S}$ \textgreater 0. Depending on $\bf P$ direction, two distinct states emerge: FE$_\text{u}$AM with [$\bf P$ \textgreater 0, $\bf {P_S}$ \textgreater 0] and FE$_\text{d}$AM with [$\bf P$ \textless 0, $\bf {P_S}$ \textless 0]. An electric field, $\varepsilon$, can switch between FE$_\text{u}$AM and FE$_\text{d}$AM states, enabling electric control of $\bf {P_S}$.		 
	}
	\label{Figure1}
\end{figure*}

According to spin group theory~\cite{Spingroup_SongPRX, Spingroup_LiuPRX, Spingroup_FangPRX, chen2025unconventional}, realizing AM requires that the magnetic sublattices be related by $R$ rather than by direct $t$ or inversion symmetry~\cite{Vsmejkal2022:beyond}. In typical FE systems, although inversion symmetry is inherently broken, the magnetic sublattices are usually connected by $t$ symmetry. This leads to conventional ferroelectric antiferromagnets (FEAFM), characterized by the spin group $[C_2||t]$, as illustrated in Figure 1(a). This incompatibility, where FE states typically preserve $t$ symmetry connecting AFM sublattices, whereas AM demands $R$ symmetry, severely restricts the number of viable FEAM, as reflected by the fact that only two candidates are identified in the MAGNDATA database.

Overcoming this symmetry conflict to realize FEAM requires introducing $R$ without compromising ferroelectricity. We propose achieving this by modifying the local environment of magnetic atoms through targeted lattice distortions such as Jahn–Teller effects, Peierls-like dimerization, and external-stimuli-induced structure deformation~\cite{Li2021:phase}. These distortions break $t$ without disrupting the FE order when they occur perpendicular to $\bf P$, leading to the emergence of $M$, a specific form of $R$, that connects the magnetic sublattices. This transformation changes the spin group from $[C_2||t]$ to $[C_2||M]$, resulting in AM with a finite $\bf {P_S}$, as shown in Figure 1(b). Remarkably, the sign of $\bf {P_S}$ is strongly coupled to the direction of $\bf P$. As the system transitions between two opposite FE configurations, FE$_\text{u}$AM and FE$_\text{d}$AM, the $\bf {P_S}$ reverses accordingly. This robust magnetoelectric coupling allows for non-volatile electric control of spin polarization without altering the N\'{e}el vector, offering a promising pathway for low-power spintronic and multiferroic applications.

\begin{figure}[t!]
	\centering
	\vspace{0.2cm}
	\includegraphics*[width=0.46\textwidth]{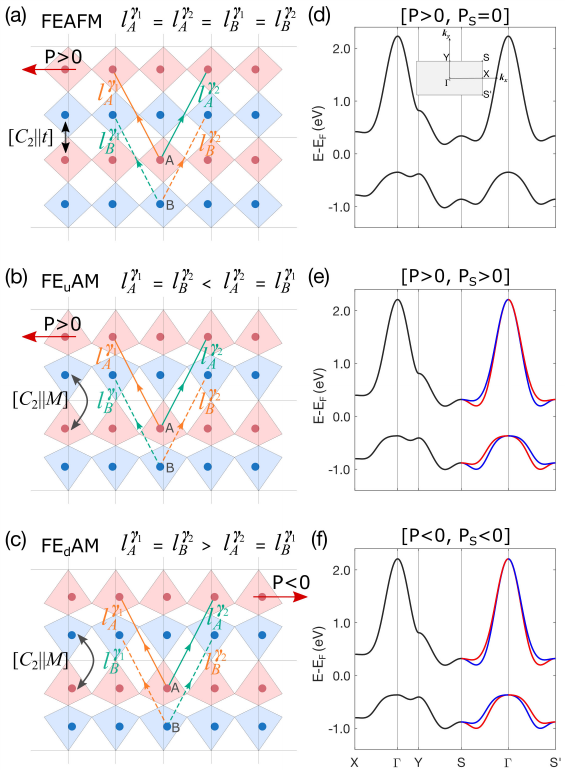}
	\vspace{-0.2cm}
	\caption{ TB model for FEAM. Schematic of a 2D rectangular AFM lattice with FE configurations and lattice distortions. Atomic sites with opposite spins (red and blue dots) form magnetic sublattices $A$ and $B$. Displacements of atoms from their centrosymmetric positions induce a net $\bf P$. \textbf{(a)} Undistorted structure with $[C_{2} \Vert t]$ symmetry gives FEAFM. \textbf{(b)} and \textbf{(c)} Distorted structure holds $[C_{2} \Vert M]$ symmetry, resulting in FE$_\text{u}$AM and FE$_\text{d}$AM, respectively. The solid (dashed) arrows indicate 4NN hopping vectors $\gamma_{1,2}$ in the $A$($B$) sublattice with hopping strengths $l_{\text{A}}^{\gamma_{1,2}}$ ($l_{\text{B}}^{\gamma_{1,2}}$). The other two hopping vectors ($\gamma_{3,4}$ = -$\gamma_{1,2}$) are not shown as the hopping strengths along opposite directions are identical ($l_{\text{A,B}}^{\gamma_{3,4}}$ = $l_{\text{A,B}}^{\gamma_{1,2}}$). \textbf{(d)}-\textbf{(f)} display the band structures calculated using TB model, where the 4NN hopping parameters are \textbf{(d)} $l_{\text{A}}^{\gamma_{1,2}}$ = $l_{\text{B}}^{\gamma_{1,2}}$ = 0.04 eV; \textbf{(e)} $l_{\text{A}}^{\gamma_{1}}$ = $l_{\text{B}}^{\gamma_{2}}$ = 0.02 eV, $l_{\text{A}}^{\gamma_{2}}$ = $l_{\text{B}}^{\gamma_{1}}$ = 0.06 eV; \textbf{(f)} $l_{\text{A}}^{\gamma_{1}}$ = $l_{\text{B}}^{\gamma_{2}}$ = 0.06 eV, $l_{\text{A}}^{\gamma_{2}}$ = $l_{\text{B}}^{\gamma_{1}}$ = 0.02 eV respectively. The inset in \textbf{(d)} illustrates the Brillouin zone and high-symmetry points. For more details about TB parameters, see Supplemental Note~S1. 
	}
	\label{Figure2}
\end{figure} 

To further reveal the mechanism of our proposed FEAM and its $\bf P$-controlled spin reversal, we construct an effective TB model based on a general 2D rectangular lattice with AFM order as in Figure 2. The model incorporates up to fourth-nearest-neighbor hopping, enabling to describe the influence of $\bf P$ and structure distortions. The Hamiltonian takes the form:

\begin{equation}
	\begin{aligned}
		H &= \sum_{i, j} \left( f_i^{\textbf{$\eta$}_j} c_i^\dagger c_{i+\textbf{$\eta$}_j} + g_i^{\textbf{$\kappa$}_j} c_i^\dagger c_{i+\textbf{$\kappa$}_j} + h_i^{\textbf{$\delta$}_j} c_i^\dagger c_{i+\textbf{$\delta$}_j} + l_i^{\textbf{$\gamma$}_j} c_i^\dagger c_{i+\textbf{$\gamma$}_j} \right) \\ &+ \text{H.C.}  + M_{A,B}\sum_{i\in A,B}c_i^\dagger \sigma_{z} c_{i}.
	\end{aligned}
\end{equation}
\noindent Here, $c_i^\dagger$ and $c_i$ represent the electron creation and annihilation operators at site $i$, respectively, and ${\sigma}$ denotes Pauli matrix. The parameters $f_i^{{\eta}_j}$, $g_i^{{\kappa}_j}$, $h_i^{{\delta}_j}$, and $l_i^{{\gamma}_j}$ describe the electron hopping between site $i$ and its first- (NN), second- (2NN), third- (3NN), and fourth-nearest neighbors (4NN), connected by the vectors ${\eta}_j$, ${\kappa}_j$, ${\delta}_j$, and ${\gamma}_j$, respectively. The AFM exchange field is characterized by $M_{A,B}$ on sublattices $A$ and $B$, with $M_{A} = -M_{B}$.

AM arises from the inequivalence of intra-spin hoppings between magnetic sublattices~\cite{Vsmejkal2022:beyond, Duan2025PRL}. Our TB model reveals that the NN, 2NN and 3NN hoppings do not contribute to AM, as they fail to generate the required spin inequivalence, even in the presence of lattice distortions, as disscused in Supplemental Note S1. In contrast, 4NN hoppings are sensitive to the lattice distortion and play a critical role in enabling AM. In the undistorted structure [Figure 2(a)], the spin group remains $[C_{2} \Vert t]$, ensuring sublattice equivalence ($l_{\text{A}}^{\gamma_{1-4}}$ = $l_{\text{B}}^{\gamma_{1-4}}$), resulting in spin-degenerate AFM [Figure 2(d)]. Upon introducing lattice distortions that lower the symmetry from $[C_{2} \Vert t]$ to $[C_{2} \Vert M]$, the 4NN hoppings become sublattice-inequivalent, i.e., $l_{\text{A}}^{\gamma_{1,3}}$ = $l_{\text{B}}^{\gamma_{2,4}}$ $\textless$ $l_{\text{A}}^{\gamma_{2,4}}$ = $l_{\text{B}}^{\gamma_{1,3}}$ [Figure 2(b)], giving rise to FEAM, as confirmed by the spin-polarized bands in Figure 2(e). Notably, reversing $\bf P$ interchanges the hopping parameters as $l_{\text{A}}^{\gamma_{1,3}}$ = $l_{\text{B}}^{\gamma_{2,4}}$ $\textgreater$ $l_{\text{A}}^{\gamma_{2,4}}$ = $l_{\text{B}}^{\gamma_{1,3}}$ Figure 2(c), and thus reverses the $\bf {P_S}$ as in Figure 2(f).

These TB model results offer critical insight into the underlying mechanism of our FEAM design. They clearly show how specific lattice distortions can preserve ferroelectricity while breaking translation symmetry, thereby inducing the spin inequivalence necessary for realizing AM. Notably, the model also reveals that reversing $\bf {P}$ naturally interchanges the hopping amplitudes between spin sublattices, leading to a $\bf {P_S}$ reversal. This makes FEAM a compelling candidate for next-generation spintronic devices. Furthermore, the required lattice distortions can arise from intrinsic mechanisms such as Jahn–Teller effects and Peierls-like dimerization, or be externally controlled via temperature, pressure, or chemical doping~\cite{Li2021:phase}. By harnessing such structure distortions to manipulate magnetic and spin properties, our model not only establishes a foundational framework for FEAM but also provides a versatile strategy for designing and exploring its potential in functional device applications.

\begin{figure*}[t!]
	\centering
	\vspace{0.2cm}
	\includegraphics*[width=0.98\textwidth]{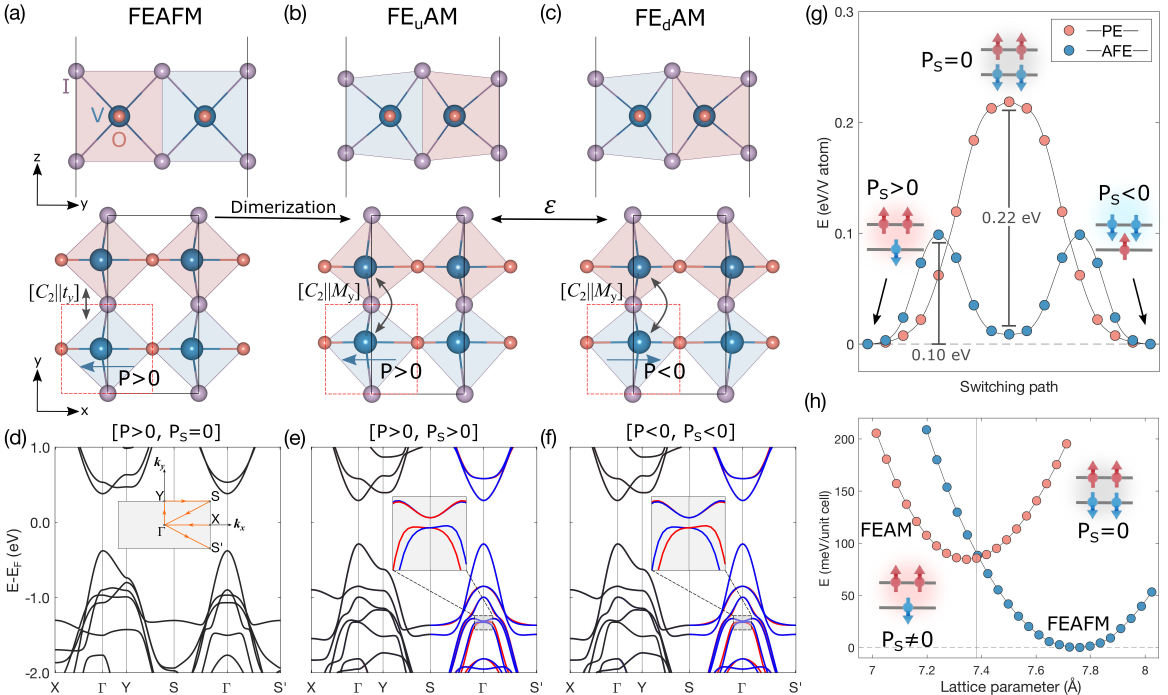}
	\vspace{-0.2cm}
	\caption{ FEAM in 2D VOI$_2$ monolayers. \textbf{(a)} Structure of undistorted monolayer VOI$_2$ with $\bf P$ \textgreater 0. \textbf{(b)} and \textbf{(c)} Distorted configurations with $\bf P$ \textgreater 0 and $\bf P$ \textless 0, respectively. Red and blue polyhedra indicate opposite spin sublattices. \textbf{(d)}-\textbf{(f)} Calculated bands for \textbf{(a)}-\textbf{(c)}, where the black, red, and blue denote the spin-degenerate, -up, and -down bands, respectively. The inset in \textbf{(d)} illustrates the Brillouin zone with its high-symmetry points. \textbf{(g)} Calculated kinetic pathways for the phase transition between FE$_\text{u}$AM and FE$_\text{d}$AM through paraelectric (PE) and antiferroelectric (AFE) intermediate states, demonstrating $\bf {P_S}$ reversal. \textbf{(h)} Lattice-constant-dependent energies of both undistorted and distorted monolayer VOI$_2$, revealing strain-tunable phase transitions between FEAFM and FEAM phases.
	}
	\label{Figure3}
\end{figure*}

We now transition from the theoretical design principles to their realization in practical materials. Based on our symmetry analysis and model construction, three key conditions are required to realize FEAM: (1) antiferromagnetic order, (2) ferroelectricity, and (3) $R$-related symmetry instead of direct $t$ symmetry connecting the magnetic sublattices. These criteria offer a universal framework for the discovery and design of FEAM materials. Guided by this principle, we identify a family of 2D vanadium oxyhalides and sulfide halides (VOX$_2$ and VSX$_2$, with X = Cl, Br, I)~\cite{tan2019two,ai2019intrinsic,xu2020electric, zhang2021peierls} as promising FEAM candidates. Monolayer VOX$_2$ compounds have been extensively studied as 2D multiferroics that violate the conventional $d$$^0$ rule, exhibiting strong polarizations and magnetoelectric coupling~\cite{tan2019two}. In this work, we focus on monolayer VOI$_2$ to demonstrate the realization of FEAM, while results for other members of the family are provided in the Supporting Information. VOI$_2$ exhibits two structural phases-- an undistorted phase [Figure 3(a)] and a distorted phase [Figure 3(b)]~\cite{tan2019two, ai2019intrinsic, xu2020electric, zhang2021peierls}-- which correspond directly to the model structures in Figure 2(a) and Figure 2(b), respectively. Our energy calculations confirm that VOI$_2$ adopts an antiferromagnetic ground state Figure 3(b) as reported in the previous work~\cite{zhang2021peierls}, with detailed comparisons of various magnetic configurations presented in Supporting Information Figure~S3 and Table S1. The monolayer VOI$_2$ structure comprises edge-sharing VO$_2$I$_4$ octahedra, where oxygen atoms are connected along the $x$-axis and iodine atoms along the $y$-axis. As shown in Figure 3(a), the FE phase emerges when V ions shift from their centrosymmetric positions within the octahedra along the $x$-direction, resulting in spontaneous $\bf P$. This symmetry breaking is driven by a pseudo Jahn–Teller effect arising from hybridization between the unoccupied V-3$d$ orbitals and occupied O-2$p$ orbitals~\cite{tan2019two}. Notably, a Peierls transition occurs along the $y$-axis within the undistorted phase, inducing V–V dimerization, which further stabilizes the distorted phase~\cite{zhang2021peierls} in Figure 3(b). This distinctive dimerization, experimentally observed in NbOCl$_2$~\cite{guo2023ultrathin} and MoOCl$_2$~\cite{wang2020fermi}, provides a perfect material platform for our model. 

We now examine the spin bands of monolayer VOI$_2$ using the exchange operation approach~\cite{noda2016momentum}, a well‐established method for studying momentum‐dependent spin splitting in AM~\cite{Duan2025PRL, okugawa2018:weakly}. Ignoring magnetism, the undistorted VOI$_2$ possesses the symmetries of $\{$$E$, $t_y$, $M_y$, $M_z$, $C_{2x}$$\}$. When AFM order is considered, its symmetries reduce to $\{$$E$, $M_z$$\}$. The missing symmetries $\{$$t_y$, $M_y$, $C_{2x}$$\}$ then act as exchange operations that link the magnetic sublattices, with $t_y$ in particular enforcing spin‐degenerate bands throughout the Brillouin zone [Figure 3(d)]. By contrast, in the distorted phase, the V–V dimerization breaks the $t_y$ symmetry, leaving $\{$$M_y$, $C_{2x}$$\}$ as the dominant exchange operations, which leads to the emergence of AM, as shown in the momentum-dependent spin splitting bands [Figure 3(e)]. The spin degeneracy along the $\Gamma$--X direction originates from the $M_y$ or $C_{2x}$ symmetry, while along the $\Gamma$--Y direction it arises from a combination of these symmetries with inversion symmetry $I$. However, for momentum along the S--$\Gamma$--S$^\prime$ direction, no symmetry operation can map them back onto themselves, resulting in spin splitting exclusively along this direction. Remarkably, when the $\bf P$ is switched between opposites FE states [Figure 3(b) and (c)], the sign of the $\bf {P_S}$ simultaneously reverses [Figure 3(e) and (f)], demonstrating a robust magnetoelectric effect in 2D VOI$_2$. 

To assess the feasibility and robustness of $\bf P$ controlled $\bf{P_S}$ switching, we evaluated the transition energy barriers between FE$_\text{u}$AM and FE$_\text{d}$AM states. We consider both the paraelectric (PE) phase, where V atoms remain undisplaced along $x$-axis, and the AFE phase, where V atoms exhibit opposite displacement, as intermediate states [Figure~3(g)]. The calculated barriers via PE and AFE states are 0.22 and 0.10 eV per V atom, respectively, comparable to other 2D multiferroic materials ~\cite{hu2019progress}. Our symmetry analysis and band structure calculations reveal that both the PE and AFE states exhibit conventional AFM behavior, i.e., [$\bf P$ = 0, $\bf {P_S}$ = 0], as shown in Supplemental Figure~S4. Therefore, during the $\bf {P}$ switching, the $\bf {P_S}$ not only undergoes the sign reversal, but also on and off toggling.

To understand the difference between the undistorted and distorted VOI$_2$ structure, we compare their relaxed lattice constants and find only a slight 0.4~$\AA$ difference along the $y$-axis [Figure~3(h)]. Applying tensile strain increases V–V distances, potentially suppressing Peierls distortion and driving the system toward the FEAFM phase. This structural evolution aligns with a clear transition from $\bf {P_S}$ $\neq$ 0 to $\bf {P_S}$ = 0. These results demonstrate that strain engineering in VOI$_2$ can serve as an effective strategy for controlling lattice distortions, particularly for achieving $t$ symmetry breaking.

The range of our proposed 2D FEAM candidates can be greatly extended by replacing the certain elements by its same group elements. This is confirmed by our calculations showing that FEAM properties maintain in other vanadium oxyhalides (VOCl$_2$, VOBr$_2$) and sulfide halides (VSX$_2$, X = Cl, Br, I), as discussed in Supplemental Figures~S5 and S6. Besides the Peierls-induced dimerization in monolayer VOI$_2$, alternative mechanisms for breaking $t$ symmetry in FEAFM systems, such as Jahn-Teller distortions (alternating long and short specific bonds)~\cite{Gu2025PRL} and rotations of magnetic lattice units~\cite{Smejkal2024arxiv}, can also induce FEAM. Fundamentally, these mechanisms all achieve spin splitting through inequivalent intra-spin hoppings, while spin reversal originates from the interchanges of hopping amplitudes between spin sublattices in two opposite ferroelectric states. 

Our theoretical framework and effective model provide a unified description of these diverse mechanisms, highlighting their general applicability. Beyond the single-phase materials proposed here, 2D FEAM can also be realized using alternative approaches such as stacking or sliding in van der Waals heterostructures~\cite{Sun2024a, sun2025proposing, zhu2025sliding}. Our single-phase strategy relies on spontaneous symmetry breaking intrinsic to the material, driven by ionic displacements, thereby enhancing experimental feasibility and potential device integration. In contrast, layer stacking/sliding strategy achieves FEAM through engineered magnetic ordering coupled with interlayer charge inhomogeneity~\cite{Sun2024a, sun2025proposing, zhu2025sliding}. Although these approaches differ in their microscopic origins of ferroelectricity, they share identical symmetry requirements, demonstrating complementary routes toward realizing our FEAM design principle.

\begin{figure}[t!]
	\centering
	\vspace{0.2cm}
	\includegraphics*[width=0.46\textwidth]{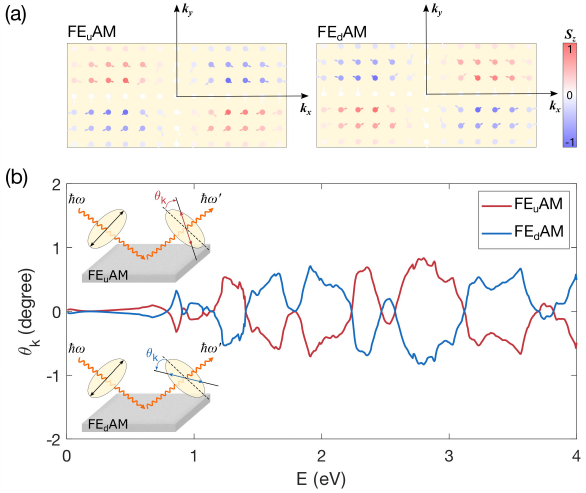}
	\vspace{-0.2cm}
	\caption{ Probing $\bf {P_S}$ reversal in FEAM. \textbf{(a)} Spin textures for FE$_\text{u}$AM and FE$_\text{d}$AM states in monolayer VOI$_2$, where in-plane spin components are indicated by arrows and out-of-plane components are represented by the color scale. \textbf{(b)} Calculated MOKE signals, characterized by the Kerr angle $\theta_k$. The inset schematically illustrates the MOKE measurement setup, with orange arrows showing the incident ($\hbar$$\omega$) and reflected ($\hbar$$\omega^\prime$) light directions and black double-headed arrows depicting the polarization directions.
	}
	\label{Figure4}
\end{figure}

Our proposed FEAM and its $\bf P$-reversed $\bf {P_S}$ can be experimentally distinguished through angle-resolved photoemission spectroscopy measurements~\cite{krempasky2024altermagnetic,jiang2025metallic,zhu2024:MnTe2}. Additionally, orientation-constrained magneto-transport measurements can also serve as an effective identification method~\cite{SongAMNature2025,vsmejkal2020crystal,shao2021spin}. Beyond the two methods discussed above, we propose an optical approach to probe spin reversal. The calculated spin texture of VOI$_2$ exhibits a distinct $d$-wave character, and the opposite spin textures observed in the FE$_\text{u}$AM and FE$_\text{d}$AM states further confirm the electric-field-switchable spin splitting [Figure~4(a)]. This reversal is further supported by MOKE measurements, which detect changes in the polarization of reflected light from magnetized surfaces, providing direct insights into magnetic properties~\cite{sivadas2016gate, ding2023magneto, feng2024layer, sun2024stacking}. Figure~4(b) shows a prominent Kerr signal in FE$_\text{u}$AM state due to anisotropic optical conductivity from time-reversal symmetry breaking. Crucially, the Kerr angle reverses correspondingly with the polarization-dependent reversal of spin splitting in the FE$_\text{d}$AM phase.

In summary, our study establishes a universal, symmetry-guided framework for realizing 2D FEAM, addressing a significant challenge in integrating ferroelectricity with altermagnetism. By combining TB models with first-principles calculations, we demonstrate that carefully engineered lattice distortions can reconcile the symmetry requirements of both orders and induce the spin inequivalence essential for altermagnetic behavior. The identification of vanadium oxyhalides and sulfide halides as promising candidates not only validates our approach but also offers experimentally accessible platforms for testing these concepts. Furthermore, leveraging their 2D nature, our proposed FEAM can be integrated into van der Waals heterostructures, opening opportunities for proximity-induced phenomena and multifunctional device architectures~\cite{geim2013van,Zutic2019,zhou2023asymmetry,zhou2022fusion}. These advances underscore the urgent need for experimental realization and application-driven studies to fully unlock the potential of 2D FEAM in next-generation, low-power spintronic and multiferroic devices.

\vspace{5mm}

\begin{acknowledgements} 
This work is supported by the National Natural Science Foundation of China (12474155, 12447163, and 11904250), the Zhejiang Provincial Natural Science Foundation of China (LR25A040001), and U.S. DOE Office of Science, Basic Energy Sciences Grant No. DE-SC0004890 (I.Z.). The computational resources for this research were provided by the High Performance Computing Platform at the Eastern Institute of Technology, Ningbo. Z.Z., X.D., and J.Z. contributed equally to this work.

\end{acknowledgements} 
\bibliography{reference}

\end{document}